\documentclass[aps,prx,twocolumn,showpacs]{revtex4-1}
\usepackage{bm}
\usepackage{mathrsfs}
\usepackage{amsmath}
\usepackage{amssymb}
\usepackage{graphicx}
\usepackage{amsfonts}
\usepackage{amsthm}
\usepackage{color}
\usepackage{dcolumn}
\usepackage{txfonts}
\usepackage[caption=false]{subfig}
\usepackage[T1]{fontenc}

\begin{document}

\title{Super-Poissonian Squeezed Light in the Deep Strong Regime of the Quantum Rabi Model}
\author{Chon-Fai Kam}
\email{Email: dubussygauss@gmail.com}
\author{Xuedong Hu}
\email{Corresponding author. Email: xhu@buffalo.edu}
\affiliation{Department of Physics, University at Buffalo, SUNY, Buffalo, New York 14260, USA}

\begin{abstract}
By analytically solving the quantum Rabi model, we investigate the photonic properties of its ground eigenstate. In particular, we find that in the deep strong coupling regime, where the coupling strength $g$ exceeds the mode frequency $\omega$, the photonic state is effectively squeezed in one of its quadratures. The squeezing reaches its maximum at the curve corresponding to the quantum phase transition of the quantum Rabi system, and decreases rapidly on both sides of the phase transition. Notably, for $g/\omega\approx 3$, which is experimentally testable in existing trapped-ion platforms, the achievable squeezing parameter can reach approximately $r\approx 0.8$. Intriguingly, the photonic state is squeezed while its number distribution follows a super-Poissonian distribution, with the largest deviation from Poissonian behavior occurring at the phase transition between the normal and superradiant phases. In other words, the ground state of the quantum Rabi model contains super-Poissonian quantum squeezed photons.
\end{abstract}

\maketitle
\textit{Introduction}.--- The coupled system consisting of a two-level system (TLS) and a bosonic field, such as an electromagnetic or phononic field, is ubiquitous in the quantum mechanical description of nature \cite{leggett1987dynamics}. A particularly interesting case in the study of atomic physics and cavity quantum electrodynamics (cavity QED) arises when the bosonic field is confined to a single mode. When both the TLS and the bosonic field are quantized, the system is fully described by the quantum Rabi model \cite{larson2007dynamics}. Historically, it has been believed that the quantum Rabi model, due to the absence of a second conserved quantity besides energy, could not be solved analytically. As a result, various approximations have been used to address the problem in specific limits. For instance, in conventional cavity QED experiments, where a single atom interacts with an optical cavity, the ratio of the atom-field interaction strength $g$ to the frequency of the single-mode radiation field $\omega$ is typically very small, with $g/\omega\approx 10^{-6}$. In this regime, the atom-cavity interaction can be effectively described by the Jaynes-Cummings model, under conditions of near resonance and weak coupling \cite{haroche2006exploring}.

Over the past two decades, in superconducting circuits \cite{langford2017experimentally} and hybrid systems \cite{stockklauser2017strong}, strong coupling limit of $g > \gamma$, where $\gamma$ is the system decoherence rate and is usually much smaller than cavity frequency $\gamma \ll \omega$, has been easily reached, and coupling strengths for such systems have been continually growing. For instance, for a superconducting flux qubit inductively coupled to an LC resonator, the ultrastrong coupling regime with $0.1<g/\omega<1$ is achieved \cite{stassi2020scalable}. In this regime, the Jaynes-Cummings model has become insufficient, and the full quantum Rabi model need to be employed. Indeed, recent experimental implementations of the quantum Rabi model, utilizing superconducting qubits, meta-materials, cold atoms, and most recently trapped ions, have achieved a $g/\omega$ ratio as high as $6.5$ \cite{dareau2018observation, lv2018quantum, yoshihara2017superconducting, bayer2017terahertz, forn2019ultrastrong, cai2021observation, koch2023quantum}, reaching into the so-called deep strong limit.  Solving the full quantum Rabi model is imperative in gaining accurate mathematical description and deep physical insights into these systems.

In 2011, it was demonstrated \cite{braak2011integrability} that the quantum Rabi model is exactly solvable, marking it the first non-integrable system \cite{berry1977level, batchelor2015integrability, xie2017quantum, eckle2019models, eckle2017generalization} that can be solved analytically. The solution takes advantage of the discrete $Z_2$ symmetry of the quantum Rabi model \cite{braak2011integrability} that arises from parity conservation, which allows the Hilbert space to be divided into two invariant subspaces \cite{casanova2010deep}. The problem can thus be formulated in the form of a two-component wave function governed by two coupled first-order ordinary differential equations \cite{xie2017quantum, bargmann1961hilbert, segal1963mathematical, zhong2013analytical, maciejewski2014full, ronveaux1995heun}. One interesting feature of the quantum Rabi model is that it exhibits a quantum phase transition from a normal phase to a superradiant phase \cite{cai2021observation} at the limit when the ratio between the atomic transition frequency $\Delta$ and the cavity field frequency $\omega$ approaches infinity $\Delta/\omega\rightarrow \infty$ \cite{zheng2023observation}.  At this limit, the quantum phase transition is governed by an effective dimensionless coupling strength $\lambda \equiv g\sqrt{2/\omega\Delta} = g/\omega \times \sqrt{2\omega/\Delta}$. The renormalized ground state energy undergoes a second-order phase transition at $\lambda=1$, characterized by a discontinuity in its second-order derivative at the critical value $\lambda_c=1$ \cite{puebla2017probing}. For $\lambda<1$, the ground state is in the normal phase, where the mean photon number remains approximately constant. In contrast, for $\lambda>1$, it enters the superradiant phase, characterized by a mean photon number proportional to $\Delta/\omega$ \cite{cai2021observation}.  This is in contrast with the Dicke model \cite{dicke1954coherence} of $N$ two-level systems uniformly coupled to a quantized single-mode cavity field, where the Dicke superradiant phase transition occurs at the thermodynamics limit as $N\rightarrow \infty$ \cite{hepp1973superradiant,wang1973phase}. 

Prior investigations \cite{hwang2015quantum} into the quantum Rabi model show that low-energy eigenstates exhibit squeezing in their photon number distributions \cite{walls1983squeezed, zhang1990coherent} in both the perturbative regime $(g/\omega\rightarrow 0)$ and the adiabatic limit $(g/\omega\rightarrow \infty)$. As $\Delta/\omega\rightarrow \infty$, the scaling behavior of squeezing suggested by the effective Hamiltonians of the Rabi model implies that significant squeezing should occur in the low-energy eigenstates in the deep strong coupling regime, even for a finite ratio of $\Delta/\omega$ \cite{hwang2015quantum}.  Experimental evidence has also indicated nonclassical photon statistics in these states, as shown by the presence of a negative region in the associated Wigner distribution under specific conditions \cite{zheng2023observation}. However, a systematic and comprehensive study of photonic properties of the quantum Rabi eigenstates has not yet been performed.

In this study, we rigorously evaluate the photonic properties of the ground state of the quantum Rabi model at finite $\Delta/\omega$ ratio, focusing on expectation values of the quadratures, correlations, and modified fluctuations. Our results indicate that, in the deep strong coupling regime—where the coupling strength surpasses the mode frequency—the ground state exhibits pronounced squeezing near the quantum phase transition. These nonclassical photon states correspond to the standard squeezed states \cite{walls1983squeezed, zhang1990coherent, kam2023coherent} in the normal phase, but they differ substantially from the standard squeezed states in the superradiant phase. Consequently, the uncertainty product is not always equal to 1/2 for all values of the dimensionless coupling strength $\lambda$, e.g., $\Delta x(\lambda)\Delta p(\lambda)>1/2$ for $\lambda\geq 1$. Quantitatively, we predict that in the deep strong coupling regime with $g/\omega\approx 3$, a value experimentally demonstrated in trapped ions, the highest attainable effective squeezing parameter could reach around $r\approx 0.8$. 

Furthermore, contrary to the common intuition that quantum states tend to be sub-Poissonian, we find that the photonic distribution in the ground state of the quantum Rabi model is super-Poissonian in both the ultrastrong and deep strong coupling regimes, for any coupling strength, regardless of whether the system is in the normal or superradiant phase. Our results provide not only rigorous mathematical descriptions of the photon statistics in the quantum Rabi model, but also practical implications and valuable insights. For instance, the significantly altered photonic properties of a cavity mode could have profound effects on cavity-related quantum information processing, such as enhancing qubit-cavity coupling and improving qubit readout fidelity.

\textit{The Model}.--- The Hamiltonian for the quantum Rabi model is given by
\begin{equation}\label{Rabi}
    H = \Delta\sigma_z +\omega a^\dagger a +g\sigma_x(a^\dagger+a),
\end{equation}
where $a$ and $a^\dagger$ are the annihilation and creation operators for a single bosonic mode with frequency $\omega$, $\sigma_x$ and $\sigma_z$ are the Pauli matrices for a qubit with level splitting 2$\Delta$, and $g$ is the interaction strength between the qubit and the bosonic mode. Without loss of generality, we choose the frequency of the bosonic mode to be $\omega=1$.  

\begin{figure}[tbp]
	\subfloat[$g<1$ \label{sfig:1a}]{%
	\includegraphics[width=0.65\columnwidth]{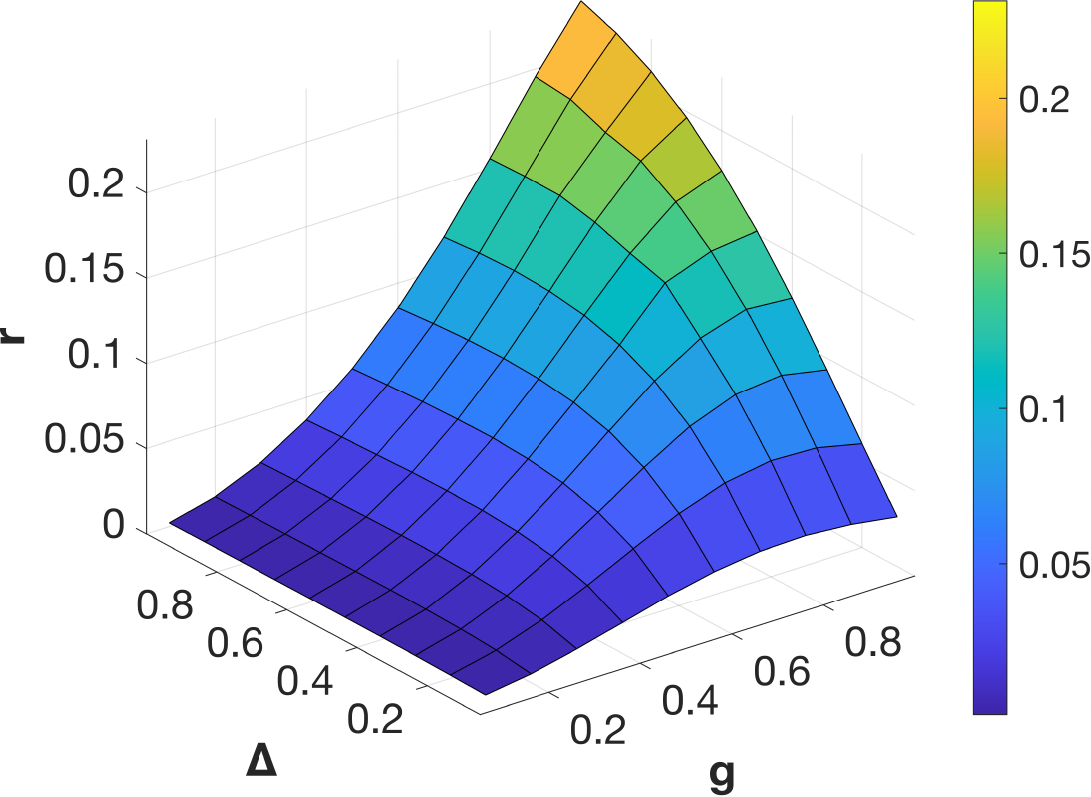}%
	}\hfill
	 \subfloat[$g>1$ \label{sfig:1b}]{%
	\includegraphics[width=0.65\columnwidth]{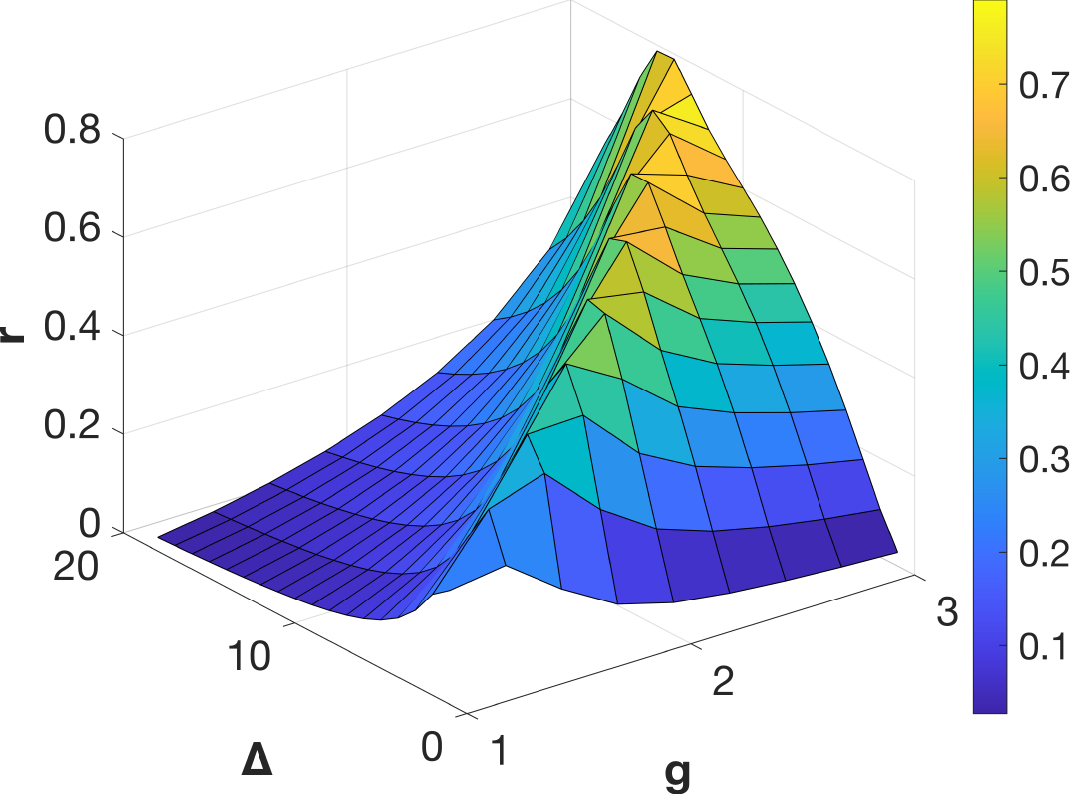}%
	}
\caption{Schematic of the squeezing parameter $r$ for the lowest odd parity eigenstate $|\psi_{0,-}\rangle$ with respect to the level-splitting $\Delta$ and the coupling strength $g$. The bosonic mode frequency is fixed at $\omega=1$, while the coupling strength $g$ is chosen to be smaller than 1 in the upper panel and larger than 1 in the lower panel.}
\label{S1}
\end{figure}

The quantum Rabi Hamiltonian has a discrete $Z_2$ symmetry: $[\Pi,H]=0$, where $\Pi\equiv e^{i\pi a^\dagger a}\sigma_z$ is the parity operator. Together with energy conservation, it allows the quantum Rabi model to be analytically solved via the Segal-Bargmann representation \cite{hall2013quantum}. Unlike the familiar Schrödinger representation, which describes quantum states as wave functions in position space, the Segal-Bargmann space offers a holomorphic representation of these states through complex entire functions. After mapping onto the Segal-Bargmann space, the eigenfunctions for the quantum Rabi model take the two-component form $\psi_{m,\sigma}(z) \equiv (\phi^{(1)}_{m,\sigma}(z),\phi^{(2)}_{m,\sigma}(z))^\top$. Here $\sigma\in\{+,-\}$ serves as a discrete label for the parity, $m\in \mathbb{N}^+$ represents the quantum number for the discrete energy levels of the photonic states within the even and odd parity subspace $H_\sigma$, and $z$ is a complex number parametrizing the phase space in the Segal-Bargmann representation, similar to the coordinate $q$ of the position space in the Schrödinger representation. Each wavefunction components have two equivalent expressions, which are given by \cite{xie2017quantum}
\begin{subequations}
\begin{align}
    \phi^{(1)}_{m,\sigma}(z) &\equiv\sigma e^{-gz}\sum_{n=0}^\infty J_n(x_m)(z+g)^n,\nonumber\\
 			        &\equiv e^{gz}\sum_{n=0}^\infty K_n(x_m)(g-z)^n,\\
    \phi^{(2)}_{m,\sigma}(z) &\equiv\sigma e^{-gz}\sum_{n=0}^\infty K_n(x_m)(z+g)^n,\nonumber\\
				&\equiv e^{gz}\sum_{n=0}^\infty J_n(x_m)(g-z)^n,
\end{align}
\end{subequations}
where $J_n(x) \equiv \frac{\Delta}{(x-n)} K_n(x)$, and $K_n(x)$ are determined recursively by \cite{braak2011integrability}
\begin{subequations}
\begin{gather}
    nK_n(x)=f_{n-1}(x)K_{n-1}(x)-K_{n-2}(x),\\
    f_n(x)\equiv 2g +\frac{1}{2g}\left(n-x+\frac{\Delta^2}{x-n}\right),
\end{gather}
\end{subequations}
where the initial values of $K_n(x)$ are fixed to be $K_{-1}(x)\equiv 0$ and $K_0(x)=1$.  Lastly, $x_m\equiv E_m+g^2$ are the renormalized eigenvalues which are zeros of the spectral function
\begin{equation}
	G_\pm(x_m)\equiv\sum_{n=0}^\infty [K_n(x_m)\mp J_n(x_m)]g^n=0,
\end{equation}

\begin{figure}[tbp]
	\subfloat[Mean photon number $\langle n\rangle$ in $|\Delta,g,\pm\rangle$ for $|\psi_{0,-}\rangle$ in the ultrastrong regime with $0.1<g<1$ \label{sfig:4a}]{%
	\includegraphics[width=0.65\columnwidth]{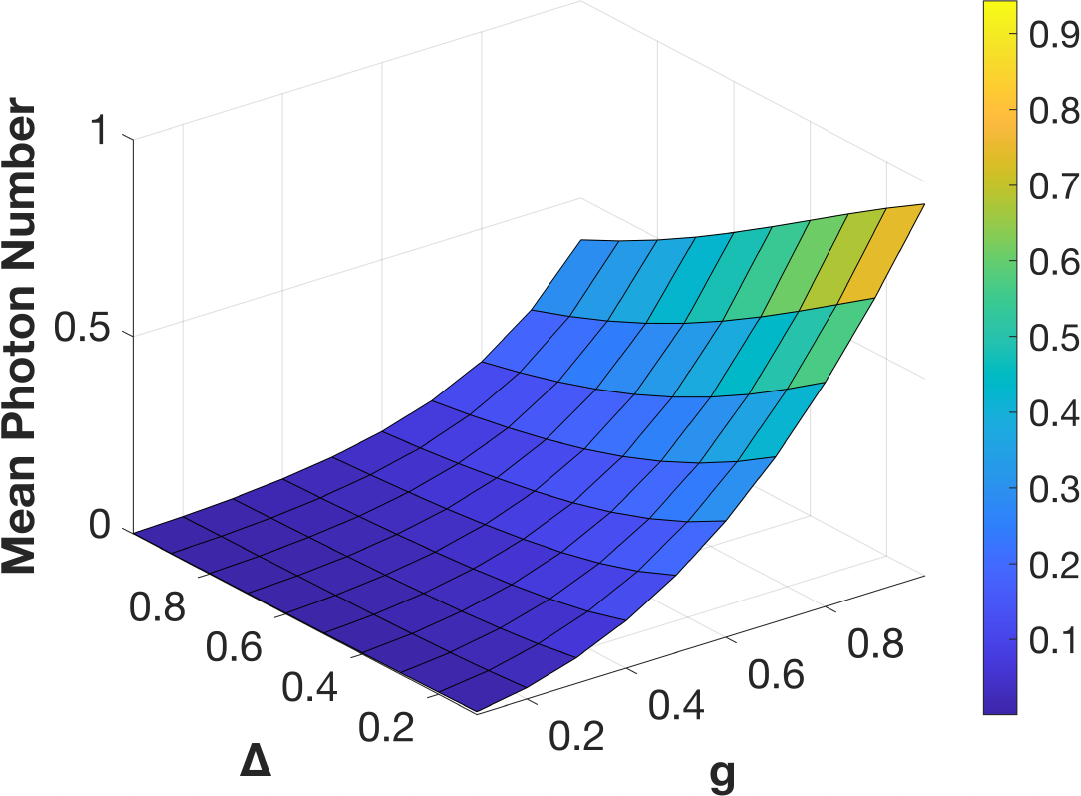}%
	}\hfill
	\subfloat[Mean photon number $\langle n\rangle$ in $|\Delta,g,\pm\rangle$ for $|\psi_{0,-}\rangle$ in the deep strong regime with $1<g<3$ \label{sfig:4b}]{%
	\includegraphics[width=0.65\columnwidth]{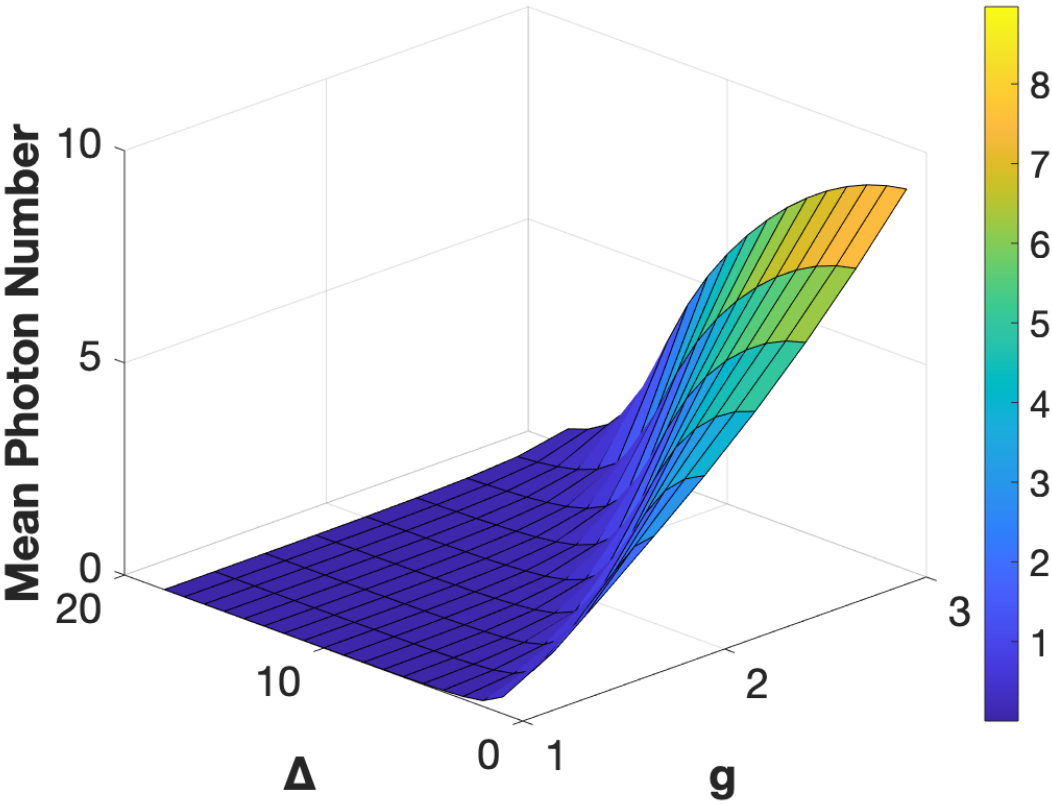}%
	}\hfill
	\subfloat[Mean $x$-quadrature in $|\Delta,g,-\rangle$ for $|\psi_{0,-}\rangle$ in the ultrastrong regime with $0.1<g<1$ \label{sfig:4c}]{%
	\includegraphics[width=0.65\columnwidth]{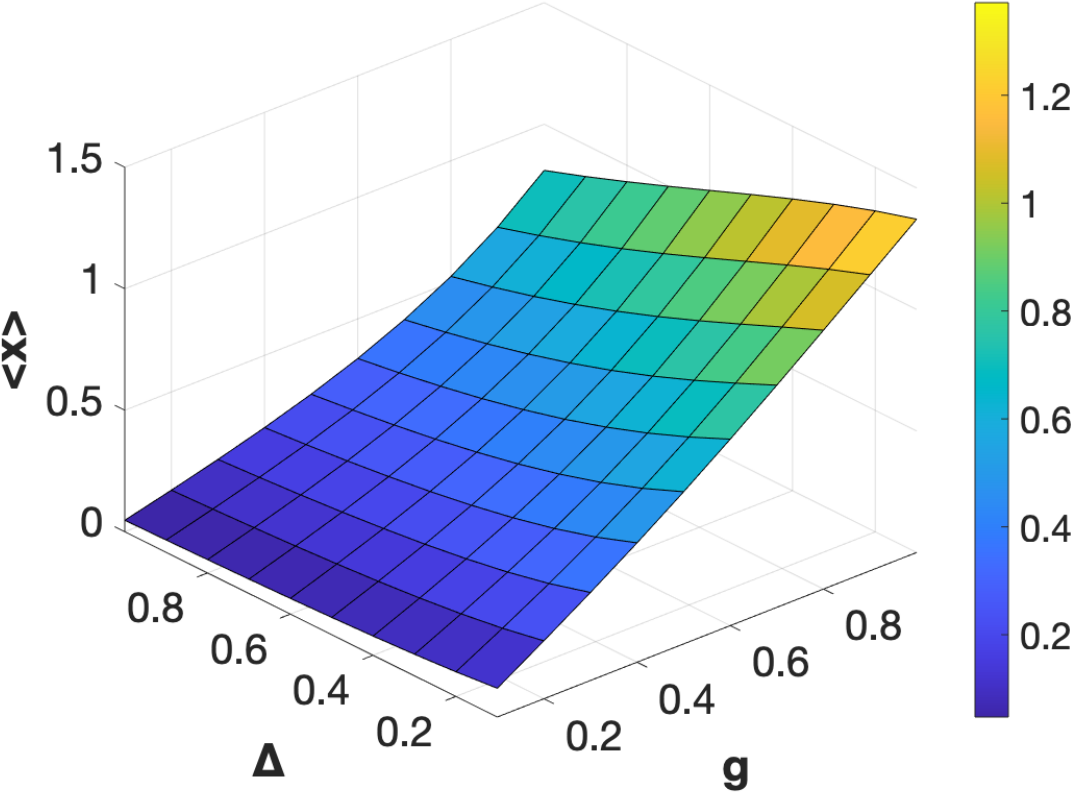}%
	}\hfill
	 \subfloat[Mean $x$-quadrature in $|\Delta,g,-\rangle$ for $|\psi_{0,-}\rangle$ in the deep strong regime with $1<g<3$ \label{sfig:4d}]{%
	\includegraphics[width=0.65\columnwidth]{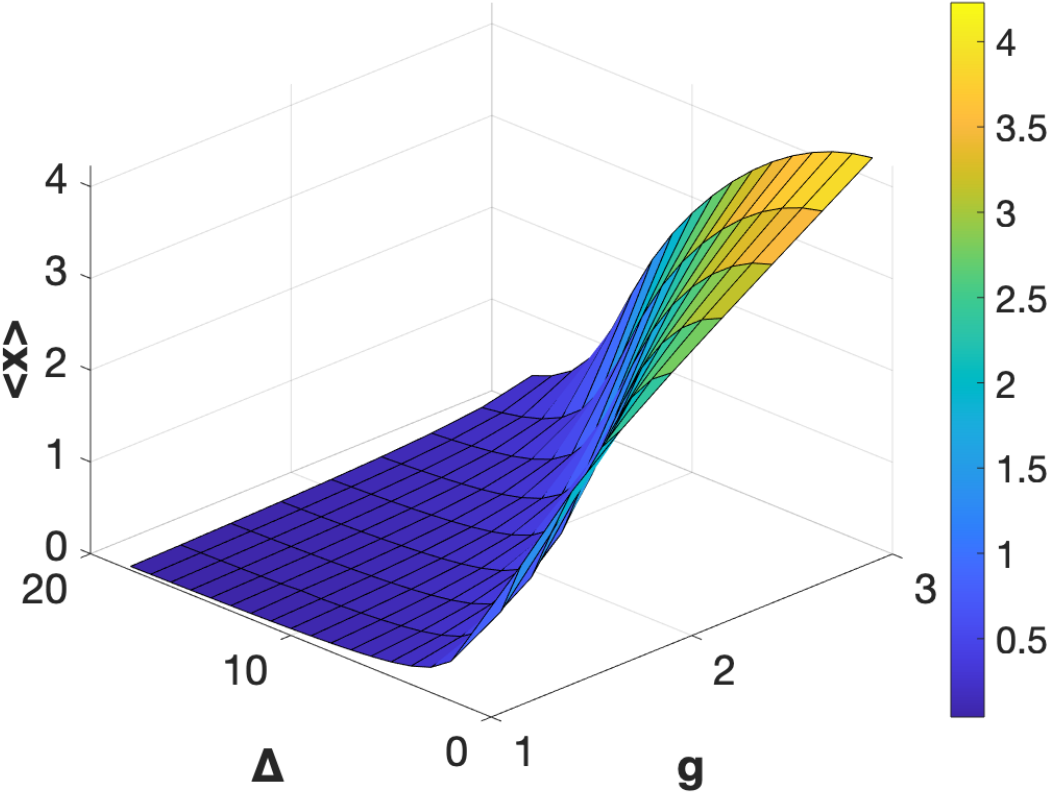}%
	}
\caption{Upper two panels: mean photon number in the photonic state $|\Delta,g,-\rangle$ (or $|\Delta,g,+\rangle$) for the lowest odd parity eigenstate $|\psi_{0,-}\rangle$ with respect to the level-splitting $\Delta$ and the coupling strength $g$. Lower two panels: mean $x$-quadrature in the photonic state $|\Delta,g,-\rangle$ for $|\psi_{0,-}\rangle$ with respect to the different $\Delta$ and $g$.}
\label{fig:Fig4}
\end{figure}

In the Hilbert space, the wavefunction components $\phi^{(1)}_{m,\sigma}(z)$ and $\phi^{(2)}_{m,\sigma}(z)$, labeled by the zeros $x_{m,\sigma}$ of the spectral function $G_\sigma(x)$, have the form $ |\Delta,g,+\rangle\equiv \phi^{(1)}_{m,\sigma}(a^\dagger)|0\rangle$ and $|\Delta,g,-\rangle\equiv \phi^{(2)}_{m,\sigma}(a^\dagger)|0\rangle$. The eigenstates of the quantum Rabi model associated with eigen-energy $E_{m,\sigma}$ is thus a cat-like entangled state
\begin{equation} 
	|\psi_{m,\sigma}\rangle\equiv |\Delta,g,+\rangle\otimes|+\rangle + |\Delta,g,-\rangle\otimes|-\rangle \,,
\end{equation} 
where $|\pm\rangle\equiv (|\uparrow\rangle+|\downarrow\rangle)/\sqrt{2}$. The photonic states $|\Delta,g,\pm\rangle$ can be equivalently expressed in terms of the eigenstates of a shifted harmonic oscillator, whose orthonormal eigenstates are $|\alpha,n\rangle \equiv D(\alpha)|n\rangle$.  Using $|\alpha,n\rangle$ as basis, $|\Delta,g,\pm\rangle$ take the form
\begin{subequations}
\begin{align}
    |\Delta,g,+\rangle&=\sigma e^{\frac{1}{2}g^2}\sum_{n=0}^\infty b_n(x_{m,\sigma})|-g,n\rangle\nonumber\\
   							 &=e^{\frac{1}{2}g^2}\sum_{n=0}^\infty a_n(x_{m,\sigma})|g,n\rangle,\\
    |\Delta,g,-\rangle&=\sigma e^{\frac{1}{2}g^2}\sum_{n=0}^\infty (-1)^na_n(x_{m,\sigma})|-g,n\rangle\nonumber\\
   							 &=e^{\frac{1}{2}g^2}\sum_{n=0}^\infty (-1)^nb_n(x_{m,\sigma})|g,n\rangle \,.
\end{align}
\end{subequations}
where $a_n(x)\equiv (-1)^n\sqrt{n!}K_n(x)$, $b_n(x)\equiv \sqrt{n!}J_n(x)$. The series expansion described above serves as the basis of this work. 
While such a series expansion has been used to derive the expectation values $\langle\sigma_x(t)\rangle$ and $\langle\sigma_z(t)\rangle$ of the spin across all regimes of coupling strength \cite{wolf2012exact}, our focus here is to determine the statistical properties of the photonic states across all regimes of coupling strength.

\begin{figure}[htbp]
	\subfloat[Uncertainly in $|\Delta,g,\pm\rangle$ for $|\psi_{0,-}\rangle$ in the ultrastrong regime with $0.1<g<1$ \label{sfig:2c}]{%
	\includegraphics[width=0.65\columnwidth]{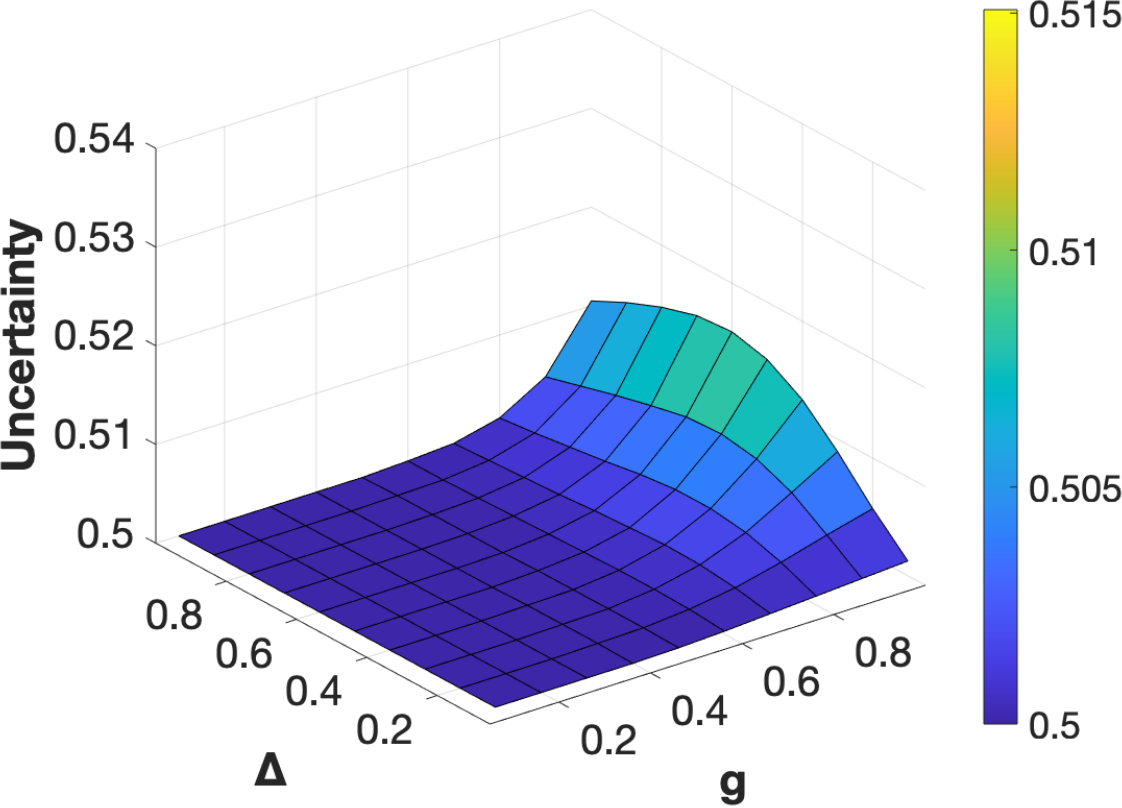}%
	}\hfill
	\subfloat[Uncertainly in $|\Delta,g,\pm\rangle$ for $|\psi_{0,-}\rangle$ in the deep strong regime with $1<g<3$ \label{sfig:2a}]{%
	\includegraphics[width=0.65\columnwidth]{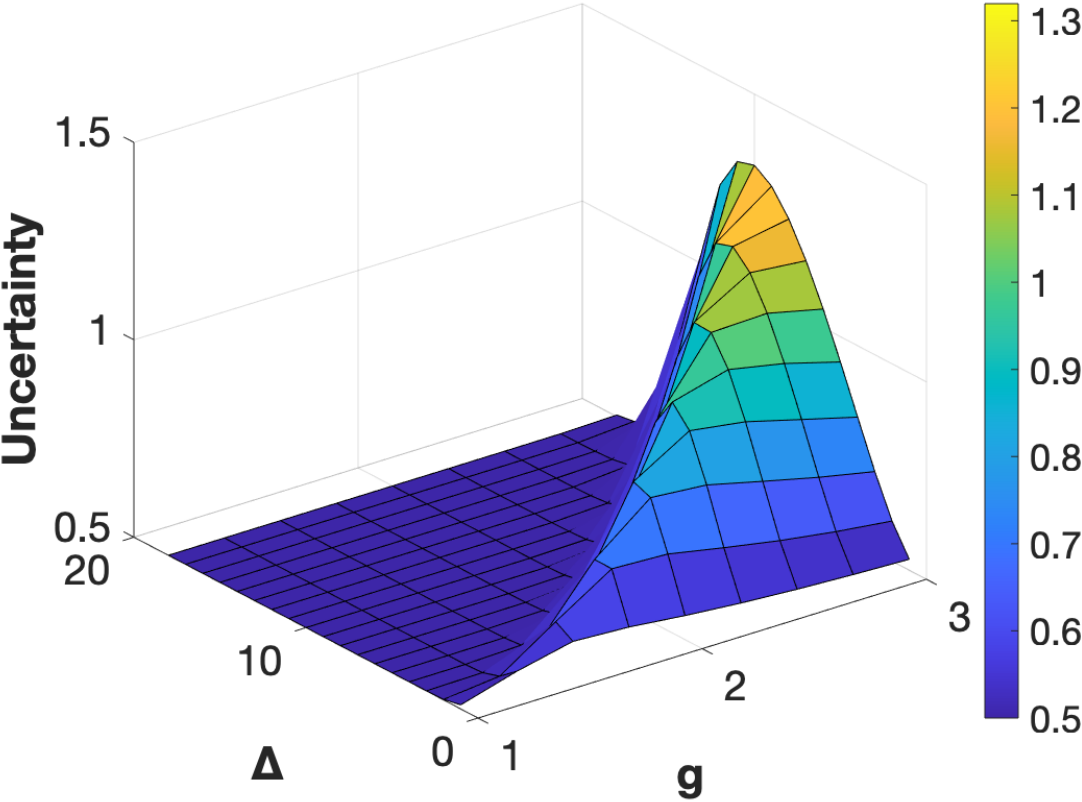}%
	}\hfill
	 \subfloat[Overlap between $|\Delta,g,+\rangle$ and $|\Delta,g,-\rangle$ for $|\psi_{0,-}\rangle$ in the ultrastrong regime with $0.1<g<1$ \label{sfig:2b}]{%
	\includegraphics[width=0.65\columnwidth]{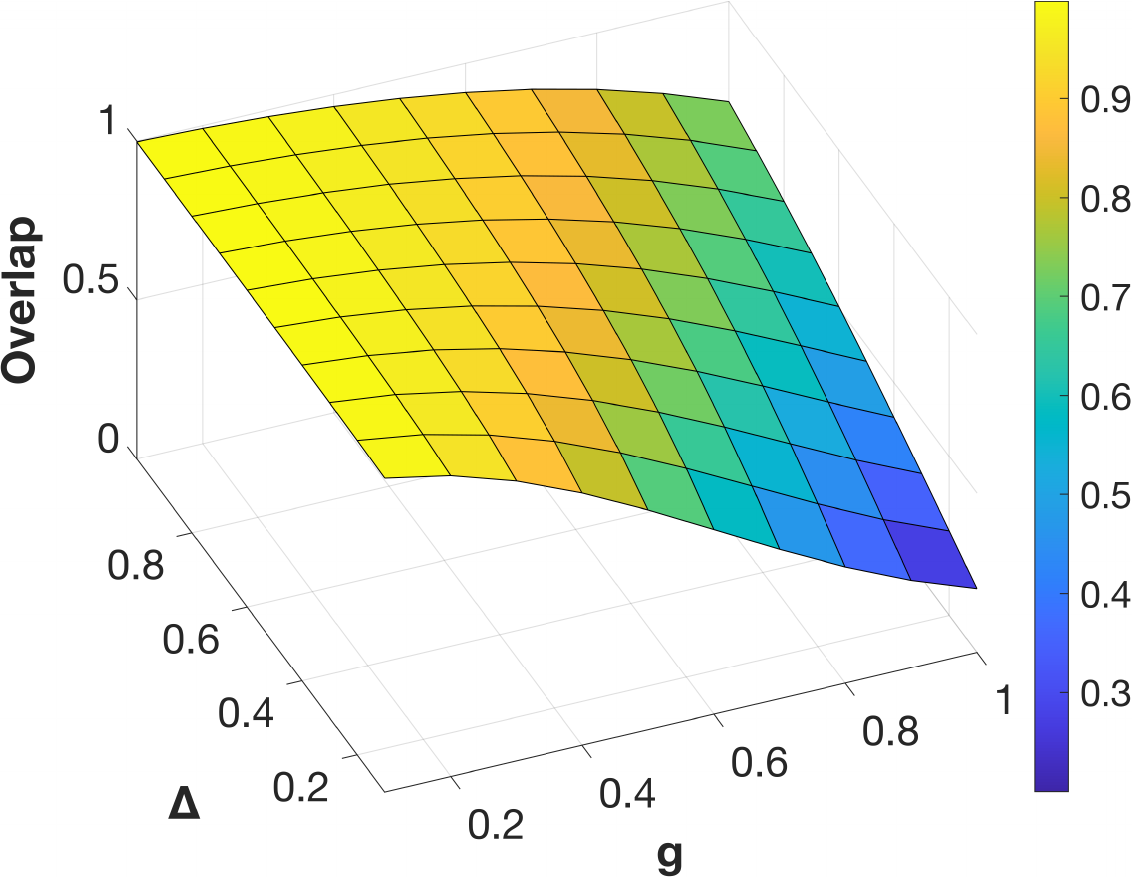}%
	}\hfill
	 \subfloat[Overlap between $|\Delta,g,+\rangle$ and $|\Delta,g,-\rangle$ for $|\psi_{0,-}\rangle$ in the deep strong regime with $1<g<3$ \label{sfig:2d}]{%
	\includegraphics[width=0.65\columnwidth]{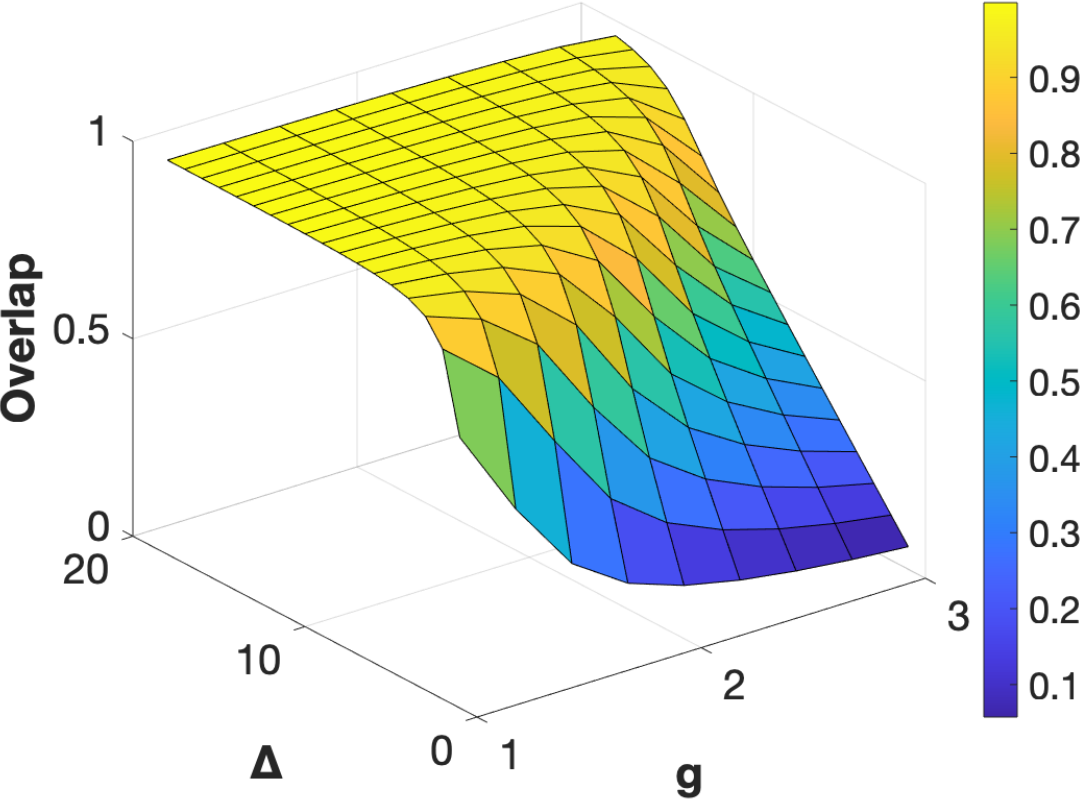}%
	}
\caption{Upper two panels: uncertainty ($\Delta x\Delta p$) in the photonic state $|\Delta,g,-\rangle$ (or $|\Delta,g,+\rangle$) for the lowest odd parity eigenstate $|\psi_{0,-}\rangle$ of the quantum Rabi model with respect to the level-splitting $\Delta$ and the coupling strength $g$. Lower two panels: the overlap between the photonic states $|\Delta,g,+\rangle$ and $|\Delta,g,-\rangle$ associated with the spin states $|+\rangle$ and $|-\rangle$ for the lowest odd parity eigenstate $|\psi_{0,-}\rangle$. The bosonic mode frequency is fixed at $\omega=1$.}
\label{fig:Fig2}
\end{figure}

\textit{Results}.--- The explicit form of the quantum Rabi eigenstates enables us to analyze the photonic statistics associated with these states. Given the known quantum phase transition in the system, we will focus on the behavior of the ground state near the phase transition, particularly in the ultrastrong and deep strong coupling regimes. In particular, we will examine the uncertainties of the quadratures $x = (a+a^\dagger)/\sqrt{2}$ and $p=(a-a^\dagger)/(i\sqrt{2})$, as well as the expectation value $\langle x\rangle$. We will explore how these quantities change across the phase transition. In addition, we also examine the photon number expectation values $\langle a^\dagger a\rangle$ and their fluctuations, and thus the associated photonic statistics. Note that the expectation $\langle p\rangle$ vanishes identically, as both $\langle a^\dagger\rangle$ and $\langle a\rangle$ are real. We calculate the photonic expectations and variances after projecting the spin states onto the $|+\rangle$ or $|-\rangle$.  When a photonic measurement is done without knowledge of the spin states, the result is the average of those projected onto the $|+\rangle$ or $|-\rangle$ states. For this type of measurement, the expectation values cancel out due to the opposite signs for the different projections, while the uncertainties and statistics remain unchanged.

We start in the ultrastrong coupling regime with $0.1 < g < 1$. we first define an effective squeezing parameter $r \equiv -\frac{1}{2}\ln(\Delta p/\Delta x)$, where $\Delta p$ and $\Delta x$ are the uncertainties in the two quadratures respectively. As illustrated in Fig.\:\ref{sfig:1a}, the upper right corner of the plot, where $\Delta = g = 1$, exhibits the highest degree of squeezing, with $r\approx 0.2$. This behavior aligns with normal expectations, as strong coupling should lead to enhanced quantum correlations, as reflected in stronger squeezing. Interestingly, however, the squeezing parameter peaks along the quantum phase transition curve, where $\Delta=2g^2$. For $g<g_c\equiv \sqrt{\Delta/2}$, increasing the coupling strength leads to larger squeezing effects; but for $g>g_c$, further increasing the coupling strength will lead to smaller squeezing effects. In the deep strong regime where $g>1$, as illustrated in Fig.\:\ref{sfig:1b}, a similar trend emerges. The squeezing parameter has a sharp peak along the curve $\Delta=2g^2$ and drops rapidly on both sides of this ridge. For $g<3$, the highest achievable squeezing parameter is approximately $0.8$. 

In the ultrastrong coupling regime, as shown in Fig.\:\ref{sfig:4a}, the mean photon number increases with a larger coupling strength. In the deep strong coupling regime with $g/\omega>1$, depicted in Fig.\:\ref{sfig:4b}, there is a continuous yet noticeable transition in the mean photon number. This abrupt change by an order of magnitude marks the onset of the superradiant phase, a key feature of the quantum phase transition in the quantum Rabi model. While the ground state contains no photons at zero coupling, it becomes populated with photons as the system enters the superradiant phase at higher coupling strengths. In contrast, the mean photon number remains essentially constant in the normal phase, corresponding to lower coupling strengths, when the energy level splitting is fixed. Similar behaviors can also be observed in the mean $x$-quadrature and the photon fluctuation, as depicted in Figs.\:\ref{sfig:4c} - \ref{sfig:4d}. The mean $x$-quadrature represents the vacuum electric field, which is qualitatively four times higher in the superradiant phase. This increase occurs with only a slight increase in the coupling strength across the phase boundary, which is remarkable. Due to the conditions $\langle a^\dagger\rangle=\langle a\rangle$, this also implies the presence of a highly coherent field. As a result, when the system enters the superradiant phase, the emission from the cavity corresponds to a strong coherent quantum light. 

Figure \ref{sfig:2c} shows that the uncertainty product $\Delta x\Delta p$ deviates from the coherent state value of $1/2$ when $g>0.4$. This is also the regime where the Jaynes-Cummings approximation of the Rabi model starts to break down. As one can see from Fig.\:\ref{sfig:2a}, the photonic state $|\Delta,g,+\rangle$ and $|\Delta,g,-\rangle$ differ from the standard squeezed state in a crucial aspect: the uncertainty $\Delta x\Delta p$ is not a constant $1/2$, but rather varies with respect to both $\Delta$ and $g$. In the normal phase, the uncertainty roughly maintains the coherent state value, whereas in the superradiant phase, the uncertainty deviates more significantly from the coherent state value. This suggests that only in the normal phase can the photonic states be approximately considered as standard squeezed states. The uncertainty peak $\Delta x\Delta p\approx 1.3$ occurs along the ridge where the quantum phase transition occurs, i.e., $\Delta\approx 2g^2$, indicating that squeezing in the $p$-quadrature comes at a greater cost of increased uncertainty in the $x$-quadrature. 

\begin{figure}[tbp]
	\subfloat[Photon fluctuation in $|\Delta,g,\pm\rangle$ for $|\psi_{0,-}\rangle$ in the ultrastrong regime with $0.1<g<1$ \label{sfig:5a}]{%
	\includegraphics[width=0.65\columnwidth]{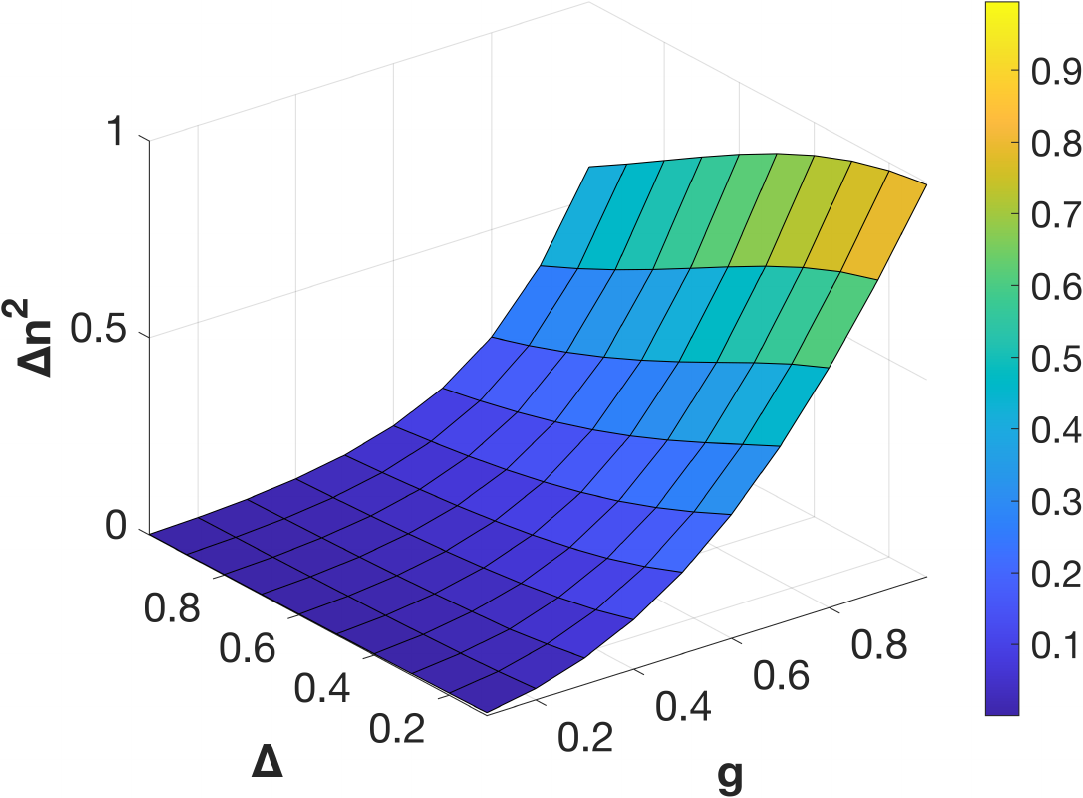}%
	}\hfill
	\subfloat[Photon fluctuation in $|\Delta,g,\pm\rangle$ for $|\psi_{0,-}\rangle$ in the deep strong regime with $1<g<3$ \label{sfig:5b}]{%
	\includegraphics[width=0.65\columnwidth]{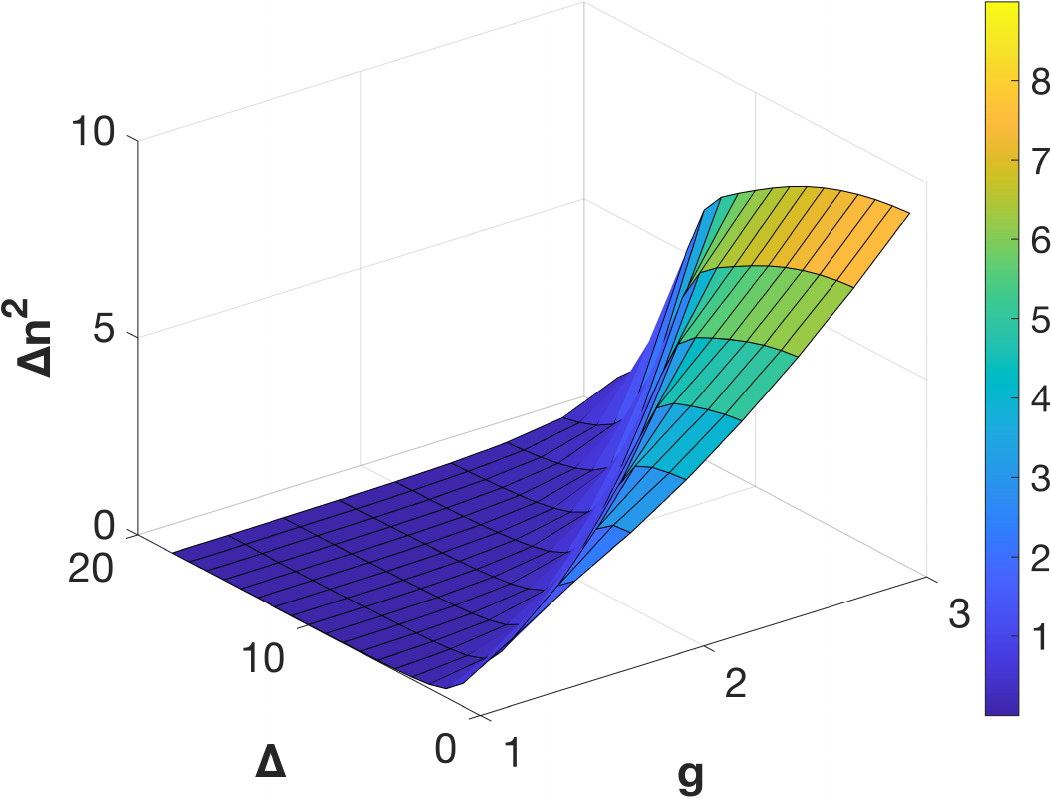}%
	}\hfill
	\subfloat[Photon statistics in $|\Delta,g,-\rangle$ for $|\psi_{0,-}\rangle$ in the ultrastrong regime with $0.1<g<1$ \label{sfig:5c}]{%
	\includegraphics[width=0.65\columnwidth]{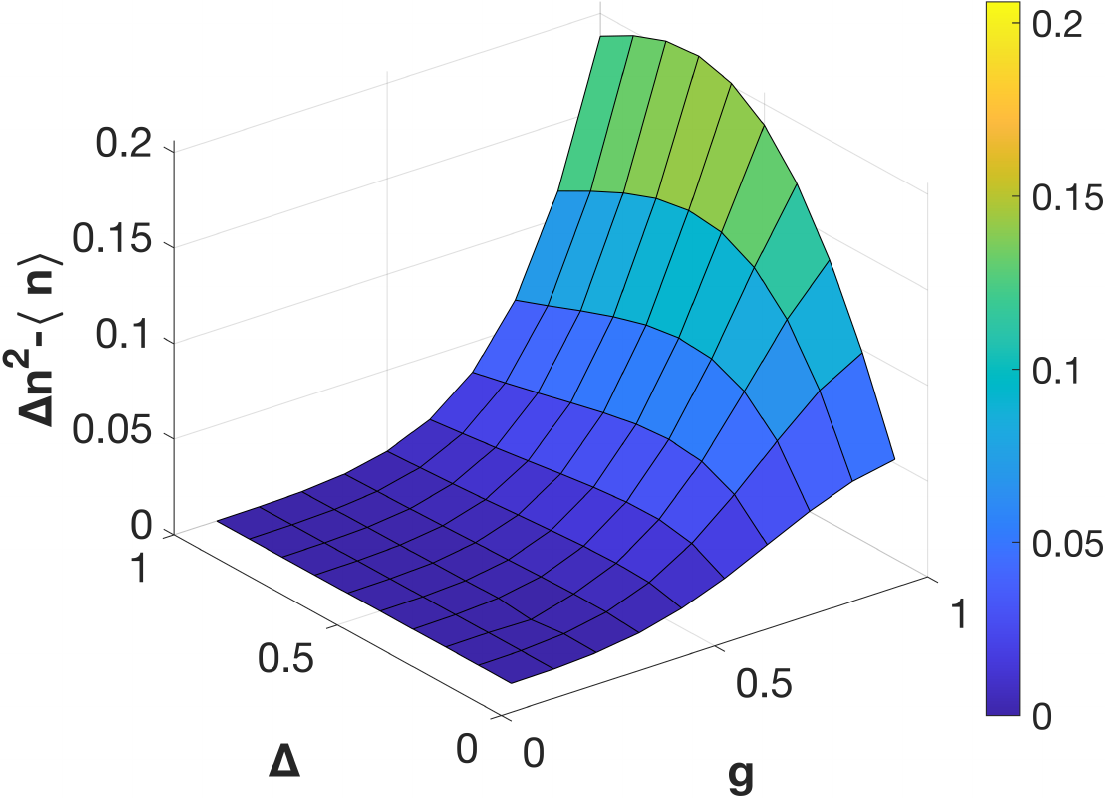}%
	}\hfill
	 \subfloat[Photon statistics in $|\Delta,g,-\rangle$ for $|\psi_{0,-}\rangle$ in the deep strong regime with $1<g<3$ \label{sfig:5d}]{%
	\includegraphics[width=0.65\columnwidth]{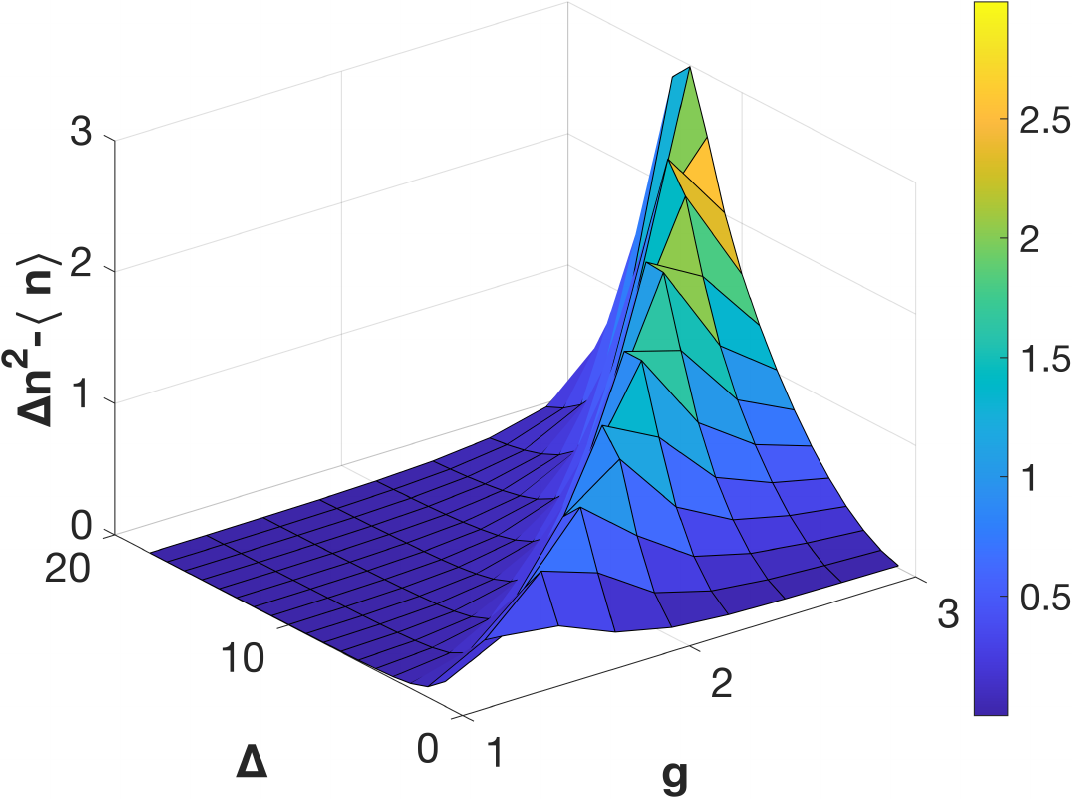}%
	}
\caption{Upper two panels: photon fluctuation $\Delta n^2$ in the photonic state $|\Delta,g,-\rangle$ (or $|\Delta,g,+\rangle$) for the lowest odd parity eigenstate $|\psi_{0,-}\rangle$ with respect to the level-splitting $\Delta$ and the coupling strength $g$. Lower two panels: photon statistics in $|\Delta,g,-\rangle$ (or $|\Delta,g,+\rangle$) for $|\psi_{0,-}\rangle$ with respect to the different $\Delta$ and $g$.}
\label{fig:Fig5}
\end{figure}

\begin{figure}[tbp]
\includegraphics[width=0.7\columnwidth]{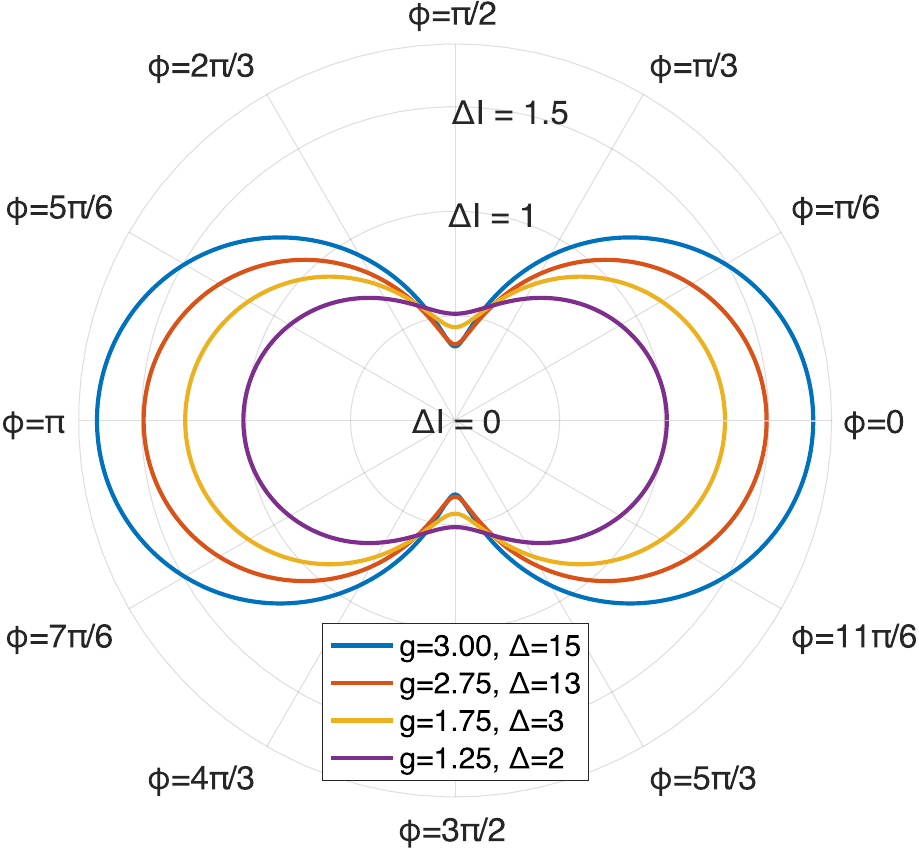}
\caption{Standard deviation $\Delta I$ of the general quadrature $I\equiv (ae^{-i\varphi}+a^\dagger e^{i\varphi})/\sqrt{2}$ equipped with an arbitrary phase $\varphi$.}
\label{S3b}
\end{figure}

\begin{figure}[tbp]
\includegraphics[width=0.55\columnwidth]{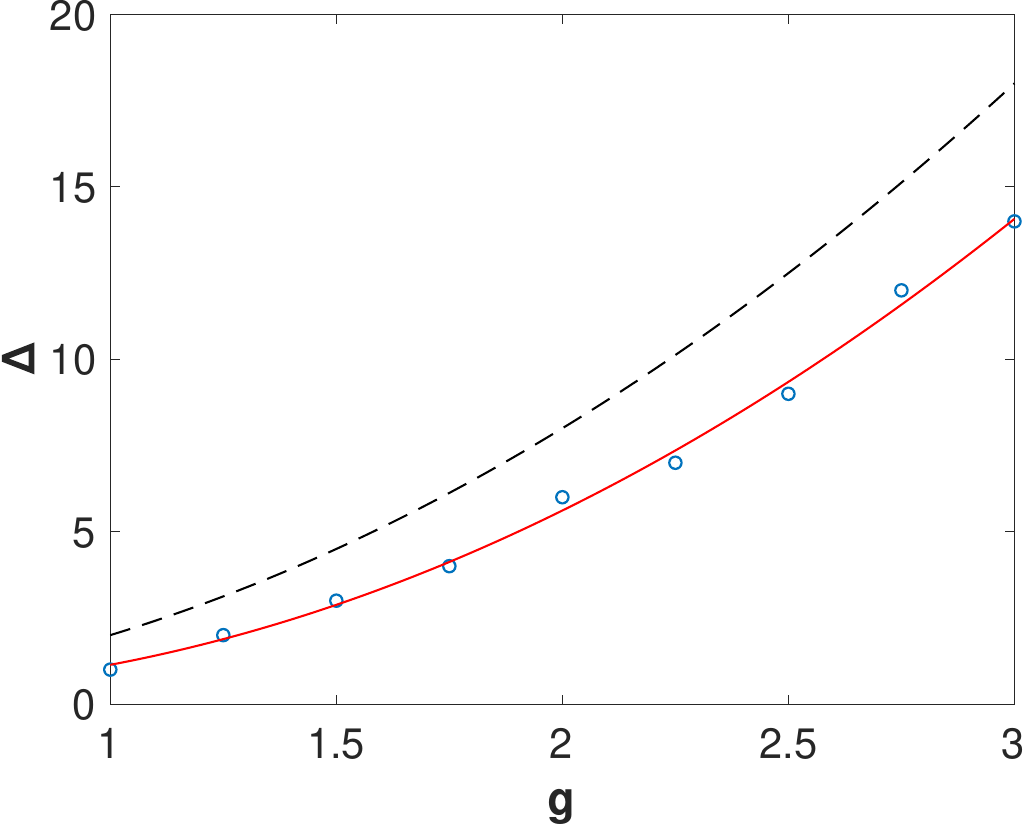}
\caption{The red solid line represents the best fit of the $\Delta-g$ curve of which the squeezing parameter is maximal. The data points are marked by blue circles. In comparison, the dark dashed line corresponds to the quantum phase transition curve $\Delta=2g^2$.}
\label{S3}
\end{figure}

Notably, as illustrated in Fig.\:\ref{sfig:2b}, for the lowest odd parity eigenstate $|\psi_{0,-}\rangle\equiv |\Delta,g,+\rangle\otimes |+\rangle + |\Delta,g,-\rangle\otimes |-\rangle $ of the quantum Rabi model, the overlap between the two photonic states $|\Delta,g,+\rangle$ and $|\Delta,g,-\rangle$ associated with the spin states $|+\rangle$ and $|-\rangle$ decreases as the coupling constant increases, while it increases with higher level-splitting $\Delta$. Particularly, in the strong coupling regime where $0.1 < g < 1$, the minimum overlap between the states $|\Delta,g,+\rangle$ and $|\Delta,g,-\rangle$ occurs at approximately $0.2$ when $g=1$ and $\Delta=0.1$. This implies that in the strong coupling regime, the two photonic states $|\Delta,g,+\rangle$ and $|\Delta,g,-\rangle$ within the lowest odd parity eigenstate $|\psi_{0,-}\rangle$ become nearly orthogonal. Consequently, the eigenstate $|\psi_{0,-}\rangle$ can be considered a Schrödinger cat state when the coupling between the two-level system and the cavity is strong and the level-splitting is small. Additionally, Fig.\:\ref{sfig:2d} illustrates that the overlap is nearly unity in the normal phase, but it rapidly decreases by an order of magnitude in the superradiant phase as the phase boundary is crossed. This indicates that the cat state picture is more relevant in the superradiant phase, where the photonic states begin to deviate from the standard squeezed states.

\begin{table*}[htbp]
\centering
\begin{tabular}{|c|c|c|c|}
\hline
& \textbf{Normal phase} & \textbf{Superradiant phase} \\ \hline
\textbf{Overlap} & \textit{Nearly Product State} & \textit{Cat State} \\ \hline
\textbf{Coherence} & \textit{Low} & \textit{High} \\ \hline
\textbf{Mean Photon Number} & \textit{Very Low} & \textit{Very High} \\ \hline
\textbf{Uncertainty}  & \textit{Nearly Squeezed State} & \textit{Deviate from Squeezed State} \\ \hline
\textbf{Squeezing} & \multicolumn{2}{c|}{\textit{Strong Squeezing at QFT}} \\ \hline
\textbf{Photon Statistics} & \multicolumn{2}{c|}{\textit{Super-Poissonian Statistics}} \\ \hline
\end{tabular}
\caption{Overview of the photon statistics of the ground state of the quantum Rabi model}
\label{tab:T1}
\end{table*}

From Figs.\:\ref{sfig:5c} - \ref{sfig:5d}, what is surprising at first glance is that the photon statistics associated with $|\Delta,g,-\rangle$ (or $|\Delta,g,+\rangle$) follow a super-Poissonian distribution. This means that the photon fluctuation is consistently larger than the mean photon number, $\Delta n^2-\langle n\rangle>0$. In traditional thinking, quantum squeezed light is often associated with sub-Poissonian statistics. Super-Poissonian statistics are typically associated with phenomena like thermal light or light from chaotic sources, where photons tend to be detected in clusters. But the logic is that the mixing of coherent states leads to super-Poissonian light, but a super-Poissonian distribution does not necessarily indicate classical light. In this case, what we have observed is a deterministically generated super-Poissonian quantum squeezed light. The quantum phase transition from the normal phase to the superradiant phase is clearly marked by a sharp, abrupt change in photon statistics. The most significant deviation from Poissonian statistics occurs along the quantum phase transition curve, while Poissonian statistics are effectively recovered far from this curve.

Interestingly, as shown in Fig.\:\ref{S3b}, the standard derivation $\Delta I$ of the general quadrature $I \equiv (ae^{-i\varphi}+a^\dagger e^{i\varphi})/\sqrt{2}$ with an arbitrary phase $\varphi$ is the \textit{spiric section} i.e., planar sections of a torus, described by the simple formula $(\Delta I)^2 = (\Delta x)^2 \cos^2 \varphi + (\Delta p)^2 \sin^2\varphi$. Thus the infinite squeezing curve corresponds to the \textit{Lemniscate of Bernoulli} (see App. II), where the squeezing curve begins to cut through the holes in the middle of the torus but stops due to the infinite squeezing. The quadrature uncertainties behave this way because the degree of anti-squeezing in the $x$-quadrature largely exceeds the degree of squeezing in the $p$-quadrature.

Finally, in Fig.\:\ref{S3}, one can see that the $\Delta - g$ curve, where the squeezing parameter reaches its maximum, does not precisely align with the curve corresponding to the quantum phase transition. A quadratic fit to the data points reveals that the $\Delta - g$ relation associated with the maximal squeezing parameter is $\Delta\approx 2g^2-1.5g+0.6$. Consequently, the actual curve lies slightly below the quantum phase transition $\Delta=2g^2$. The deviation of the actual quantum phase transition curve from the one obtained through perturbative computation indicates that the perturbative treatment may overestimate the superradiance phase. The actual area of the actual superradiance phase in the $\Delta-g$ parameter space is slightly smaller.

To summarize all the results, we present the photon statistical properties of the quantum Rabi model in Table\:\ref{tab:T1}.

\textit{Conclusion}.--- In our study, we rigorously investigate the photon statistics of the ground state in the quantum Rabi model. Even at a finite ratio $\Delta/\omega$, the ground state exhibits strong squeezing in the deep strong coupling regime where the coupling strength surpasses the mode frequency. Specifically, for $g/\omega\approx 3$, the squeezing parameter reaches approximately $0.8$. This level of squeezing holds significant potential for applications in quantum information processing, such as improving qubit readout fidelity \cite{kam2024fast}. Our findings may inspire novel approaches for designing hybrid qubit platforms aimed at fast and high-fidelity qubit readout.

In future work, it will be valuable to investigate the properties of excited-state quantum phase transitions \cite{caprio2008excited} in the quantum Rabi model. In this study, we focus on the photonic statistics of the ground state, where we observe super-Poissonian statistics instead of the expected sub-Poissonian behavior. We also identify strong photonic squeezing, particularly in the $p$-quadrature, in the deep strong coupling regime. This raises intriguing questions about whether similar unusual statistics might also emerge in the higher excited states—a topic that will be explored in future studies.

Our study also has broader conceptual implications. In the standard approach, emission from the cavity is typically analyzed using a Hanbury Brown and Twiss (HBT) interferometer via the intensity correlation function $g^{(2)}(0)$. The HBT experiment is historically important because classical wave theory can explain both Poissonian and super-Poissonian photon statistics, but it cannot account for sub-Poissonian photon statistics, a defining feature of quantum light. However, in the strong and deep-strong coupling regimes, our rigorous computations show that photon statistics alone are insufficient to distinguish whether the cavity emission is quantum or classical. This leads to the crucial question: \textit{How can one uniquely identify quantum light?}

It appears that more refined tools and experimental setups are needed to fully characterize quantum light. What about higher-order correlations? How are these refined quantum statistics connected to entanglement properties for the eigenstates? It seems that changes in the entanglement class are not captured by the HBT experiment. But how can entanglement be directly measured—not through full tomography or Bell-like experiments, but via correlations? How do internal two-photon processes influence higher-order photon correlations? What are the applications of nonclassical light sources in quantum metrology? And what about open quantum systems? These are all important future directions yet to be explored.

\begin{acknowledgements}
We acknowledge financial support by US ARO via grants W911NF1710257 and W911NF2310018. C. F. K. acknowledge financial support by the National Natural Science Foundation of China (Grant nos. 12104524).
\end{acknowledgements}

\end{document}